\documentclass[12pt]{article}

\usepackage{epsf}

\begin{document}


\newcommand\balpha{\mbox{\boldmath $\alpha$}}
\newcommand\bbeta{\mbox{\boldmath $\beta$}}
\newcommand\bgamma{\mbox{\boldmath $\gamma$}}
\newcommand\bomega{\mbox{\boldmath $\omega$}}
\newcommand\blambda{\mbox{\boldmath $\lambda$}}
\newcommand\bmu{\mbox{\boldmath $\mu$}}
\newcommand\bphi{\mbox{\boldmath $\phi$}}
\newcommand\bzeta{\mbox{\boldmath $\zeta$}}
\newcommand\bsigma{\mbox{\boldmath $\sigma$}}
\newcommand\bepsilon{\mbox{\boldmath $\epsilon$}}

\newcommand{\be}{\begin{eqnarray}}
\newcommand{\ee}{\end{eqnarray}}
\newcommand{\nn}{\nonumber}

\newcommand{\sign}{{\rm sign}\,}
\newcommand{\keff}{\kappa^{\rm eff}}
\newcommand{\zeff}{\zeta^{\rm eff}}

\newcommand{\ft}[2]{{\textstyle\frac{#1}{#2}}}
\newcommand{\eqn}[1]{(\ref{#1})}
\newcommand{\vsone}{\vspace{1cm}}
 
\begin{titlepage}

\begin{flushright}
hep-th/9911094\\
KCL-MTH-99-45\\
SWAT/244
\end{flushright}
\begin{centering}
\vspace{.2in}
{\Large {\bf Mirror Symmetry and Toric Geometry in Three Dimensional 
Gauge Theories}}\\
\vspace{.4in}
Nick Dorey$^{1}$ and David Tong$^{2}$  \\
\vspace{.4in}
$^{1}$Department of Physics, University of Wales Swansea,\\
Singleton Park, Swansea, SA2 8PP, UK\\
{\tt n.dorey@swan.ac.uk}\\
\vspace{.2in}
$^{2}$Department of Mathematics, Kings College, \\
The Strand, London, WC2R 2LS, UK\\
{\tt tong@mth.kcl.ac.uk}\\
\vspace{.6in}
{\bf Abstract} \\
\end{centering}
\vspace{0.2in}
We study three dimensional gauge theories with ${\cal N}=2$ 
supersymmetry. We show that the Coulomb branches of such theories
may be rendered compact by the dynamical generation of Chern-Simons 
terms and present a new class of mirror symmetric theories in which 
both Coulomb and Higgs branches have a natural description in terms 
of toric geometry.

\vspace{.1in}


\end{titlepage}

\section{Introduction}

Mirror symmetry of three dimensional gauge theories is an infra-red 
equivalence of two theories in which Coulomb and Higgs branches, and 
Fayet-Iliopoulos parameters and masses are exchanged. First 
discovered in ${\cal N}=4$ theories by Intriligator and Seiberg \cite{is}, 
explanations in terms of string theory dualities \cite{hw} 
as well as generalisations to both other gauge groups \cite{berk} and 
${\cal N}=2$ gauge theories \cite{berk2,ahiss} soon appeared. More recently 
it has been shown that, for a large class of three-dimensional
theories, the correspondence may be extended to all length 
scales \cite{as}. 

In this paper we present further examples of mirror theories with ${\cal N}=2$ 
supersymmetry. A novel feature of these theories is that the 
Coulomb branch may be compact and is naturally described in the 
language of toric geometry. In fact we will find that the Coulomb branch,  
which in three dimensions admits a toric action, possesses submanifolds 
on which certain cycles of the torus vanish and thereby defines a toric 
variety. Further we find pairs of theories 
for which the Coulomb (Higgs) branch of one theory and the Higgs 
(Coulomb) branch of the other are specified by the same toric data. 
A classical analysis of the Higgs branch in question 
simply yields the usual symplectic quotient construction of the 
corresponding toric variety. In contrast, 
various quantum effects which are characteristic to three dimensions
play a central role in realizing the same toric variety as  
the Coulomb branch of the mirror theory. This equivalence means, in
particular, that the two branches admit identical U(1) 
isometries with precisely the same set of fixed submanifolds. As we are
dealing with theories with only four supercharges it is not obvious 
whether the correspondence extends to the respective metrics on the two
branches. Despite this, we will find that the two metrics do in fact 
agree in cases where an explicit calculation is possible

The plan of the paper is as follows: in Section 2 we review various 
aspects of abelian gauge theories in three dimensions. In Section 3 
we present a simple self-mirror theory for which both the Higgs and 
Coulomb branch is a copy if ${\bf CP}^1$. Finally, in Section 4 we 
consider a more general abelian gauge theory and employ the language of  
toric geometry. The appendix contains a discussion of the brane realisation 
of the theory of Section 3. As this work was in preparation, \cite{ak}
appeared which considers similar theories. The connection between
three-dimensional Chern-Simons gauge theories and the related
phenomenon of mirror symmetry in 
two-dimensional Calabi-Yau $\sigma$-models has been
discussed in \cite{kog}. 

\section{Review of N=2 Gauge Theories}

Three dimensional gauge theories with ${\cal N}=2$ supersymmetry 
(4 supercharges) have been studied in detail in \cite{ahiss}, 
where a comprehensive introduction may be found. 
Here we collate some facts relevant to abelian gauge theories, with 
a $U(1)^r$ gauge group and $N$ chiral multiplets. 
To this end, we introduce 
abelian vector superfields $V_a$, $a=1,\cdots,r$, and chiral 
superfields $Q_i$, $i=1\cdots,N$, both of which are simply the dimensional 
reduction of the familiar four dimensional ${\cal N}=1$ superfields. 
The gauge kinetic terms are written most simply in terms of linear 
superfields $\Sigma_a=\epsilon^{\alpha\beta}\bar{D}_\alpha D_\beta V_a$, 
whose lowest component is a real scalar $\phi_a$ and also includes 
the $U(1)$ field strength as well as two Majorana fermions. 
The chiral multiplets each consist of a 
complex scalar $q_i$ and two further Majorana fermions. The kinetic terms 
for all fields are written as D-terms,
\be
{\cal L}_K=\int{\rm d}^4\theta\,\left[\sum_{i=1}^NQ^\dagger_i
\exp\left(2\sum_{a=1}^rR_i^aV_a\right)Q_i +\sum_{a=1}^r
\frac{1}{e_a^2}\Sigma_a^2\right]
\label{above}\ee
where $e^a$ is the $a^{\rm th}$ gauge coupling constant which has dimension 
(mass)$^{1/2}$ and $R_i^a$ are the charges of the chiral multiplets under 
each of the gauge symmetries. We assume these charges to be integers. 
Further interactions 
take the form of a superpotential, ${\cal W}$, constructed from gauge 
invariant monomials of the chiral superfields,
\be
{\cal L}_F=\int{\rm d}^2\theta\,{\cal W}(Q_i)\ +\ {\rm h.c.}
\nn\ee
In particular, if there exist two chiral superfields of opposite 
charge the usual complex mass is written is this manner. In 
three dimensions each chiral multiplet may have a further, real, 
mass parameter which cannot 
be written in terms of a superpotential and which will play an 
important role in the following discussion. It is introduced by 
weakly gauging the Cartan-subalgebra of the global flavour symmetry 
of the theory and constraining the vector multiplet scalar to a 
fixed background vacuum expectation value (VEV). The net result is that 
the exponent in \eqn{above} is replaced by,
\be
\sum_{a=1}^rR_i^aV_a \rightarrow \sum_{a=1}^rR_i^aV_a + 
2m_i\theta\theta^\dagger
\nn\ee
Notice that there are only $N-r$ independent such parameters, 
the remaining $r$ being set to zero by shifts of the vector 
multiplet scalars $\phi_a$.

Two further sets of couplings will also prove important in the story: 
Fayet-Iliopoulis (FI) parameters $\zeta_a$ which have dimension 
of mass, and dimensionless Chern-Simons (CS) 
parameters $\kappa_{ab}$. The former are incorporated in the 
usual fashion,
\be
{\cal L}_{FI} = \sum_{a=1}^r\zeta_a\int{\rm d}^4\theta\,V_a
\nn\ee
while the latter are written in terms of the linear superfield as,
\be
{\cal L}_{CS}=\sum_{a,b=1}^r\kappa_{ab}\int{\rm d}^4\theta\,\Sigma_aV_b
\nn\ee
Notice from the similarity of these two expression that the 
combinations of scalar fields $\sum_b\kappa_{ab}\phi_b$ will play 
the role of a dynamical FI parameter in the theory. This may be 
seen by examining the 
classical scalar potential, obtained by integrating out auxiliary 
fields,
\be 
U=\sum_{a=1}^r\,e_a^2\left(\sum_{i=1}^NR_i^a|q_i|^2-
\sum_{b=1}^r\phi_b\kappa_{ab}-\zeta_a\right)^2 
+\sum_{i=1}^N M_i^2|q_i|^2
+\sum_{i=1}^N\,\left|\frac{\partial{\cal W}}{\partial q_i}\right|^2
\label{U}\ee
where 
\be
M_i=\sum_{a=1}^rR^a_i\phi_a+m_i
\label{mi}\ee 
is the effective mass of the $i^{\rm th}$ chiral multiplet.
The manifold of classical supersymmetric vacua, determined by the condition 
$U=0$, depends on the parameters $\zeta_a,\kappa_{ab}$ and $m_i$. 
Let us consider situations with ${\cal W}=0$. Then there may 
be two branches of vacua: the Higgs and Coulomb branches. 
In the former, the vector multiplet scalars are set to zero, 
$\phi_a=0$, while the $q_i$'s are constrained by the vanishing of the 
first expression in \eqn{U} modulo gauge transformations. In this phase 
the gauge symmetry is generically completely broken. It is clear from 
the form of \eqn{U} that the Higgs branch does not exist for generic 
non-zero real masses. 

On the Coulomb branch, all chiral multiplet scalars are set to zero, 
$q_i=0$, while the $\phi_a$'s are  unconstrained. Again, 
from \eqn{U}, it is clear that non-zero FI or CS parameters will 
lift the Coulomb branch. When this phase exists however 
the gauge symmetry is completely unbroken and one may exchange 
each abelian gauge field for a scalar, $\sigma_a$, of period $e_a$, 
via the duality transformation 
$F^{\mu\nu}_a=\epsilon^{\mu\nu\rho}\partial_\rho\sigma_a$. The Coulomb 
branch is therefore parametrised by the VEVs of both $\phi_a$ and 
$\sigma_a$, which combine to lie in $r$ chiral multiplets, and is 
classically given by $({\bf R}\times{\bf S}^1)^r$. There exist $r$ $U(1)_J$ 
isometries of the Coulomb branch induced by constant shifts of the 
$r$ dual photons. These are preserved in the full quantum theory
\cite{ahiss}. In the following it will sometimes be 
useful to weakly gauge these symmetries. In particular, consider 
gauging the symmetry which shifts $\sigma_{a}$ by a constant and
leaves the other dual photons invariant. As explained in 
\cite{ahiss}, the lowest component of the linear
multiplet which contains the corresponding field strength is precisely
the FI parameter $\zeta_{a}$. In addition to pure Higgs and Coulomb
branches there will also exist mixed branches of vacua in which
both $q$'s and $\phi$'s have non-zero VEV. 
 
This concludes our discussion of the classical field theory. Let us 
now mention some relevant quantum effects that occur. Firstly note that in 
four dimensions a theory with the above matter content would suffer 
a gauge anomaly unless $\sum_iR_i^a=0$ for each gauge group $a$. In 
three dimensions, while there are no gauge anomalies, the theory 
may suffer from a ``parity anomaly'' \cite{red}. This manifests 
itself in the dynamical generation of CS terms due to integrating 
out chiral multiplet fermions at one-loop \cite{ahiss}. In the case 
of ${\cal W}=0$, the effective CS term is given by,
\be
\keff_{ab}=\kappa_{ab}+\ft12\sum_{i=1}^N\,R^a_iR^b_i\,\sign M_i
\label{wherekwent}\ee
The parity anomaly arises from the observation that gauge invariance 
requires $\keff_{ab}$ to be an integer and therefore, for certain matter 
content, the factor of $\ft12$ in \eqn{wherekwent} means a non-zero 
bare CS term is obligatory, breaking parity. 

A similar CS coupling is also generated for weakly gauged global 
symmetries of the type discussed earlier in this section. In 
particular, for the element of the Cartan-subalgebra of the global 
flavour group under which $q_i$ transforms with charge $+1$, with all 
other fields neutral, one has
\be
\keff_{ai}= \ft12R_{a}^{i}\,\sign M_{i} 
\nn\ee
Finally, the combined effect of all dynamically generated CS parameters may be 
interpreted as a finite renormalisation of the FI parameters, 
\be
\zeff_a= \zeta_a + \sum_{b=1}^r\keff_{ab}\phi_{b} + \sum_{i=1}^N 
\keff_{ai}m_{i}
\label{zeff}\ee
Notice that the gauge and global symmetries appear on an equal footing 
in this expression. Further, for $M_i=0$, which is the case on the Higgs 
branch, $\sign M_i$ and $\keff_{ab}$ are ill-defined. However, the 
FI parameters $\zeff_a$ are of the form $M_i\, \sign M_i$ and are 
therefore continuous at $M_i=0$. Finally, these one-loop corrections 
can be implemented in the scalar potential \eqn{U}, simply by
replacing the FI parameters and CS terms by the renormalised FI 
parameters \eqn{zeff}. 

There is one further quantum effect, discussed in \cite{ahiss}, that will 
be important for us. At the intersection of Coulomb and 
Higgs branches, certain $U(1)_J$ isometries, which are generically 
broken on the Coulomb branch, must be restored as the  
Higgs branch is invariant under such symmetries. This, in turn, 
requires the vanishing of the periods of the dualized scalars at these points 
and the Coulomb branch is to be viewed as a fibre of $T^r$ over 
$R^r$ where certain cycles of the torus shrink at the intersection 
points. This suggests that the Coulomb branch can be thought of as a 
toric variety specified in terms of toric data which encodes its 
intersections with other branches.  In the following Sections we will 
realize this idea explicitly.

\section{A Self-Mirror Example}

In this section we exhibit a simple ${\cal N}=2$ gauge theory in 
which both Higgs and Coulomb branches are 
compact, the former classically and the latter due to the dynamical 
generation of CS terms. This will also serve as an pedagogical example 
for more complicated models to be discussed in the following section. 
The theory has a single abelian gauge group with bare FI parameter $\zeta$. 
The matter content consists of two chiral multiplets, 
both transforming with charge $+1$ and with real masses 
$m_1=-m_2$. We set both superpotential and bare CS parameter to zero  so 
that the effective one-loop couplings are given by,
\be
\kappa^{\rm eff}=\ft12 \left[ \sign(\phi+m)+\sign(\phi-m)\right]
\label{kaphere}\ee
Similarly, the one-loop effective FI parameter is,
\be
\zeta^{\rm eff}=\zeta+\ft12 (\phi+m)\,\sign(\phi+m)+ \ft12 (\phi-m)
\,\sign(\phi-m)
\label{fihere}
\ee
and the scalar potential \eqn{U} thus simplifies to,
\be
U=e^2\left(|q_1|^2+|q_2|^2-\zeff\right)^2+(\phi+m)^2|q_1|^2
+(\phi-m)^2|q_2|^2
\nn\ee
Let us now examine solutions of the vacuum condition $U=0$ of this theory as a 
function of the FI and mass parameters. Without loss of generality we
may restrict our attention to masses $m\geq 0$. We will also set 
the bare CS coupling to zero. There are then five different regimes: 

i) $m=\zeta=0$: In this case there is a unique vacuum at the 
origin $\phi=q_1=q_2=0$.

ii) $m=0$, $\zeta>0$: In this case we have a 
Higgs branch of vacua given by $\phi=0$, 
and $|q_1|^2+|q_2|^2=\zeta$ modulo gauge transformations. This is simply 
Witten's gauged linear sigma model \cite{wit} with target space 
${\bf CP}^1$ of K\"ahler class $\zeta$.

iii) $m>0$, $\zeta=-m$: In this case, 
requiring $\zeff=0$ restricts the range of $\phi$ to $|\phi|\leq m$. 
Therefore the space of vacua is given by $q_1=q_2=0$ 
while $\phi$ may take any value in the interval, 
$I=\{\phi : -m\leq\phi\leq m\}$. We will discuss 
this case in more detail below.

iv) $m>0$, $\zeta>-m$: In this case there are two isolated Higgs
branch vacua. These are located at $\phi=-m$,  $|q_{1}|^{2}=m$,
$q_{2}=0$ and $\phi=m$, $|q_{2}|^{2}=m$, $q_{1}=0$ respectively.

v) $m\geq 0$, $\zeta<-m$: In this case there are two isolated vacua
on the Coulomb branch located at $q_1=q_2=0$, $\phi=\pm \zeta$.

The above vacuum structure may be reproduced by realising the theory 
as a D3-brane probe of a $(p,q)$ 5-brane web. This is done in the 
appendix, where we also discuss the vacuum structure in the 
presence of a non-zero bare CS term.

The novel feature of case iii) above 
is the restriction of the range of $\phi$ to an interval $I$. For the 
critical value of the bare FI parameter, $\zeta=-m$, 
the Coulomb branch is therefore compact. 
After dualizing the massless photon in favour of a periodic scalar
$\sigma$, the Coulomb branch can be thought
as a fibration of $S^1$ over $I$. The $U(1)_J$ 
symmetry which shifts $\sigma$ by a constant acts on the fibre over each
point. Classically the fibration is trivial and 
simply corresponds to the cylinder $I\times S^1$. However we 
will now argue that this picture is modified by quantum
effects. The key idea, based on the arguments of \cite{ahiss}, is that
the endpoints of the interval, $\phi=\pm m$, must be invariant under $U(1)_J$. 
This follows because, at these two points, the Coulomb branch 
intersects the Higgs branch of case iv) which must be invariant under 
$U(1)_{J}$. Strictly speaking, moving onto the Higgs branch requires
changing the bare FI parameter $\zeta$ away from its critical value. 
This means that one is moving to a different theory rather 
than onto another vacuum branch of the same theory. This distinction 
is irrelevant here because, as explained in the previous section, we
can promote $\zeta$ to be the scalar
component of a background linear multiplet by weakly gauging
$U(1)_{J}$. The existence of a new branch emenating from the endpoints
of the interval then hinges on whether or not this field is massless. 
At a generic point on the Coulomb branch $U(1)_{J}$ acts non-trivially
on the dual photon is therefore spontaneously broken. This means that 
the corresponding gauge multiplet which contains $\zeta$ acquires a
mass by the Higgs mechanism. Conversely, $\zeta$ only remains 
massless at points on the Coulomb branch at which $U(1)_{J}$ remains 
unbroken.  

The restoration of $U(1)_{J}$ discussed above is only possible if the $S^{1}$ 
fibre shrinks to zero size at the endpoints of the interval. Following
\cite{ahiss}, this 
effect which can be ascribed to quantum corrections which are 
unsupressed near these points. The result is a Coulomb branch with 
the topology of a two sphere. 
As mentioned above, it is natural to
combine $\phi$ and $\sigma$ to form a complex scalar which is the
lowest component of a chiral superfield. The Coulomb branch can then
be thought of as a K\"{a}hler manifold of complex dimension one. 
We therefore seek such a complex manifold which can be realized as 
an $S^1$ fibration over an interval with degenerate fibres at the
endpoints. This is precisely the data which specifies ${\bf CP}^1$ 
as a toric variety. In fact, as explained in \cite{vl}, 
the interval is just the toric diagram for ${\bf CP}^{1}$. In the following
we will see that this connection between three-dimensional Coulomb
branches and toric diagrams is a general one.   

In this simple case, one can also understand the symmetry between
Higgs and Coulomb branches by considering the original 
${\cal N}=4$ self-mirror theory of Intriligator and Seiberg \cite{is}. 
In the ${\cal N}=2$ language this consists of a $U(1)$ 
vector multiplet, a single neutral chiral multiplet, two chiral 
multiplets of charge $+1$ and two of charge $-1$. A superpotential 
couples all hypermultiplets. One may flow to the theory described 
above by gauging various global symmetries (including a subgroup of 
one the $SU(2)$ R-symmetries of the ${\cal N}=4$ algebra) and introducing 
mass terms for the unwanted chiral multiplets. The duality properties 
should be invariant under such deformations \cite{ahiss}, and the 
Higgs and Coulomb branches should again be equal in the above theory. 

Finally, one may also see the emergence of the $SU(2)$-invariant
K\"ahler metric on ${\bf CP}^{1}$ by an explicit one-loop
calculation on the Coulomb branch. Note however that, 
unlike the theory with ${\cal N}=4$ supersymmetry, one has little 
control over the K\"ahler potential and the following calculation 
is only valid at points where $e^2\ll|\phi\pm m|$ and, in particular, 
cannot be trusted at the end points of the interval. Nevertheless we 
proceed with the calculation, encouraged by the end result. 
Classically, the low-energy dynamics on the Coulomb branch 
are described by the metric,
\be 
{\rm d}s^2=H(\phi){\rm d}\phi^2+H(\phi)^{-1}{\rm d}\sigma^2
\label{met}\ee
where classically $H=1/e^2$. As well as the renormalisation of the 
FI and CS parameters discussed above, there is also a finite 
renormalisation of the gauge coupling constant,
\be
H^{\rm 1-loop}&=&\frac{1}{e^2}+\frac{\ft12}{|\phi+m|}
+\frac{\ft12}{|\phi-m|}\nn\\
&=&\frac{1}{e^2}+\frac{m}{m^2-\phi^2}
\nn\ee
where the second equality follows only when $\phi$ is restricted to the 
interval $I$. 
In the limit, $e^2\rightarrow\infty$, we thus  
find that one-loop Coulomb branch metric indeed becomes the
Fubini-Study  metric on ${\bf CP}^1$ with K\"ahler class $m$.  
Notice that, as in the original Intriligator-Seiberg model, 
the $U(1)_J$ symmetry of the 
Coulomb branch is enhanced in the infra-red to a $SU(2)$ symmetry. 
Under mirror symmetry this is exchanged with the $SU(2)$ flavour 
symmetry of the theory while the FI parameter is exchanged 
with the real mass.

To summarise, this model is self-mirror: the Coulomb and Higgs 
branches coincide if one simultaeously exchanges the 
mass and FI parameters. Specifically, the Coulomb branch exists  
only when $\zeta =0$ and is given by ${\bf CP}^1$ of Kahler class 
$m$. In constrast, the Higgs branch only exists when $m=0$ and is 
given by ${\bf CP}^1$ of Kahler class $\zeta$. 
\section{Mirror Symmetry and Toric Geometry}

In this section we discuss the more general mirror symmetric theories. 
Specifically, we will consider,
\paragraph{}
{\bf Theory A:} $U(1)^r$ gauge theory with $N$ chiral multiplets containing 
scalars $q_i$ of 
charge $R_i^a$, $i=1,\cdots,N$ and $a=1,\cdots,r$. The parameters of 
the model are bare CS couplings $\kappa_{ab}$, bare FI parameters 
$\zeta_a$ and real masses $m_i$. Notice that only $N-r$ of the 
mass parameters are independent. 
\paragraph{}
{\bf Theory B:} $U(1)^{N-r}$ gauge theory with $N$ chiral multiplets 
containing scalars $\tilde{q}_i$ 
of charge $S_i^u$, $u=1,\cdots,N-r$. The parameters are 
$\tilde{\kappa}_{uv}$, $\tilde{\zeta}_u$ and $\tilde{m}_i$, where 
$N-(N-r)=r$ of the mass parameters are independent. 
\paragraph{}
The charges of Theory A and Theory B are  constrained to satisfy,
\be
\sum_{i=1}^NR_i^aS_i^u=0\ \ \ \ \ \ \ \mbox{\rm for all}\ a\ \mbox{\rm and}\ u
\label{theone}\ee

We denote the Coulomb branch of Theory A (B) as ${\cal M}_{C}^{A\,(B)}$ 
and the Higgs branch as ${\cal M}_H^{A\,(B)}$. Notice firstly that 
$dim({\cal M}_C^{A})=dim({\cal M}_H^B)=r$ and 
$\dim({\cal M}_C^B)=dim({\cal M}_H^A)=N-r$. Similarly, the number 
of independent mass parameters of Theory A (B) is equal to the number 
of FI parameters of Theory B (A). 

Let us start with the Higgs branch of Theory B in the case where the 
masses and CS parameters vanish. This is defined as the symplectic 
quotient,
\be
{\cal M}_H^B=\mu_a^{-1}(\zeta_a)/U(1)^{N-r}
\nn\ee
where the momentum map is $\mu_a=\sum_iS^u_i|\tilde{q}_i|^2$. For 
certain values of the FI parameters (specifically, when $\zeta_a$ 
lie in the K\"ahler cone of ${\cal M}_H^B$ - see \cite{mp}), the 
classical Higgs branch is the toric 
variety\footnote{For an introduction to toric geometry for physicists, 
see section 3 of \cite{mp} and section 4 of \cite{kmv}. Our presentation 
will follow these references}, 
\be
{\cal M}_H^B=({\bf C}^{N}-F_{\Delta})/({\bf C}^{\star})^{N-r}
\nn\ee
in which ${\bf C}^N$ is parametrised by $\tilde{q}_i$, and 
$F_\Delta$ is a subset of ${\bf C}^N-{\bf C}^{\star N}$. Precisely 
which subset is determined in terms of a fan of cones, $\Delta$, as 
will be reviewed below. The action of $({\bf C}^{\star})^{N-r}$ in  
the above quotient is the complexified gauge symmetry of Theory B, 
given by,
\be
\tilde{q}_i\rightarrow \lambda^{S_i^u}\tilde{q}_i\ \ \ \ \ \ \ \ \ \ \ \ 
\ \ \ \lambda\in{\bf C}^\star\, ,\, u=1,\cdots N-r
\nn\ee
Our first task is to identify the set $F_\Delta$. This is specified 
by the charges $S_i^u$, which we may 
view as $(N-r)$ charge vectors in ${\bf Z}^N$. The first step is to 
construct $r$ vectors orthogonal to 
$S_i^u$. These are provided by the charges of Theory A, $R_i^a$, 
courtesy of equation \eqn{theone}. These charges then provide a convenient 
basis of gauge invariant polynomials which parametrise ${\cal M}_H^B$,
\be
X(k)=\prod_{i=1}^N \tilde{q}_i^{\,\langle R^i,k\rangle}\ \ \ \ \ \ 
\ \ \ \ \mbox{where}\ \langle R^i,k\rangle=\sum_{a=1}^rR^i_ak_a
\nn\ee
with $k_a\in{\bf Z}^r$. The charges, $R_i^a$, are now in turn viewed 
as $N$ vertices in ${\bf Z}^r$ and it is these which are used to 
define $\Delta$ as the collection of cones bounded by vectors from the 
origin through the vertices. Finally, the set $F_{\Delta}$ is defined 
as containing all sets $\{\tilde{q}_i=0\,:\,i\in \{i_\rho\}\,\}$ 
such that the corresponding set of vertices 
$\{R_{i}^a\,:\,i\in\{i_\rho\}\,\}$ does not lie in any single cone of 
$\Delta$. 

The toric variety defined in this manner has an action of 
${\bf C}^{\star r}$. For each vertex $R_a^i\in{\bf Z}^r$, one 
may define the action 
\be
\tilde{q}_j\rightarrow\lambda^{\delta_{ij}}\tilde{q}_j
\label{cyclebaby}\ee
which is simply the complexified global flavour symmetry of Theory B. 
The limit point, $\lambda=0$, of this symmetry is contained 
${\cal M}_H^B$ (as opposed to $F_\Delta$) as the vertices are themselves 
the integral generators of one-dimensional cones and therefore clearly 
belong to a single cone. Notice however that not all linear combinations 
of the symmetry need necessarily have limit point in ${\cal M}_H^B$. 
Nevertheless, the vertices $R_i^a$ do encode the fixed point structure of all 
abelian isometries of ${\cal M}_H^B$ and allow one to reconstruct 
the geometry of the toric variety. This is the approach 
to toric geometry discussed in \cite{vl}. To each of the vertices 
$R_i^a$, one associates a a hyperplane, $D_i$, orthogonal to the vertex, 
on which the corresponding cycle \eqn{cyclebaby} is taken to vanish, 
\be
D_i\equiv\left\{Y_a\in{\bf R}^r: \sum_{a=1}^rR_i^aY_a=C_i\right\}
\label{dual}\ee
where $C_i$ are $N$ constants. Define $\nabla$, 
a polytope of of dimension $r$, as the region of ${\bf R}^r$ 
bounded by the $N$ hyperplanes $D_i$. Importantly, by construction,  
a collection of hyperplanes only intersect if the corresponding 
limit points are not in $F_\Delta$. This ensures that the Higgs branch, 
${\cal M}_H^B$ may be viewed as a fibration of $T^r$ over 
$\nabla$ such that on the boundary $D_i$ the cycle defined 
by $R_i^a$ in equation \eqn{cyclebaby} shrinks while, on the 
intersection of $k$ boundaries, $k$ such cycles shrink. 

If the vertices of $\nabla$ lie on ${\bf Z}^r$, then ${\nabla}$ is 
said to be an integral reflexive polytope. These vertices then 
encode the information for a dual toric variety. This is 
Batyrev's construction of mirror manifolds \cite{batty}.

This concludes our discussion of the Higgs branch. The next part 
of the story is to show that the Coulomb branch of Theory A, 
${\cal M}_C^A$, corresponds to the same toric variety. ${\cal M}_C^A$ 
is parametrised by the $r$ real scalars $\phi_a$ and 
$r$ dual photons, $\sigma_a$. The key point is that the $\phi$'s are 
restricted to lie within $\nabla$ due to CS terms, while certain 
periods of the $\sigma$'s vanish on the boundaries of this space
thus giving a non-trivial fibration of $T^r$ over $\nabla$ as 
described above. To see this, take the constants $C_i$ appearing in 
\eqn{dual} to be defined by $C_i=-m_i$ for $i=1,\cdots,N$. 
Then the equation $M_i=0$, defined in \eqn{mi}, specifies the hyperplane 
$D_i$, defined in \eqn{dual}, spanned by the $\phi_a=Y_a\in{\bf R}^r$. 
Thus the region $\nabla$ is specified by a choice of $\sign(M_i)$ 
for each $i$. In order to fix this choice, consider the effective 
CS coupling \eqn{wherekwent},
\be
\keff_{ab}=\kappa_{ab}+\ft12\sum_{i=1}^N\,R_i^aR_i^b\sign\,(M_i)
\nn\ee
We see that a judicious choice of bare CS coupling $\kappa_{ab}$ 
will ensure that $\keff_{ab}$ vanishes in $\nabla$ and only 
in $\nabla$. As long as $\keff_{ab}$ vanishes, it follows from
equation (\ref{zeff}), that we may choose the bare FI parameters 
$\zeta_{a}$ in such a way that the effective FI parameters $\zeff_{a}$
vanish. In this case the part of the Coulomb branch described 
by the $\phi_a$ is precisely $\nabla$.

As the hyperplane $D_i$ is defined by $M_i=0$, one chiral
multiplet becomes massless on each component of the boundary of
$\nabla$. Thus, at least in the classical theory, 
the hyperplane $D_{i}$ is the root of a Higgs branch 
on which $q_{i}\neq 0$ and $q_{j}=0$ for all $j\neq i$. As in the
simple self-mirror example of the previous section, moving onto this
Higgs branch requires changing the FI parameters. In fact, for each
$i$, moving onto the corresponding Higgs branch requires varying the 
specific linear combination $R_{i}^{a}\zeta^{a}$ away from its
critical value. To analyse this transition, it is convenient to 
promote this linear combination of the FI parameters 
to a background superfield. After dualizing the gauge fields to periodic
scalars, this corresponds to weakly gauging a particular subgroup of the
$U(1)^{r}_{J}$ symmetry, $U(1)^{(i)}_{J}$, which shifts the
corresponding linear combination, $R_{i}^{a}\sigma_{a}$, of the 
dual photons. Following the same arguments as in Section 3, the 
existence of the new branch in the quantum theory is only consistent 
if this symmetry is unbroken. At a generic point on the Coulomb branch 
the whole of $U(1)_{J}^{r}$ is spontaneously broken. 
Hence, we predict that the subgroup $U(1)^{(i)}_{J}$ must be restored 
on the intersection between the two branches
which in turn requires that the corresponding cycle of the toric fibre
must degenerate over $D_{i}$. One may easily check that this is
exactly the same cycle as that defined in equation
(\ref{cyclebaby}). This means that Coulomb branch of Theory A 
precisely agrees with the description of ${\cal M}_B^H$ given 
above as a toric fibration of the polytope $\nabla$.  
This completes our identification of ${\cal M}_C^A$ with ${\cal
M}_H^B$.

Let us illustrate the above ideas with the  simple example of 
complex projective space. The relevant toric data for this example 
can also be found in section 4 of \cite{kmv}. We take,
\paragraph{}
{\bf Theory A:} $U(1)^{N-1}$ gauge theory with $N$ chiral multiplets 
of charge $R_i^a=\delta^a_i-\delta_i^{a+1}$. The masses are 
$m_i=-m$ for $i=1,N$ and $m_i=0$ for $i=2,\cdots,N-1$ where $m>0$. 
The bare CS parameters are
\be
\kappa_{ab}=\delta_{ab}-\ft12\delta_{a,b-1}-\ft12\delta_{a-1,b}
\label{itskagain}\ee 

{\bf Theory B:} $U(1)$ gauge theory with $N$ chiral multiplets of 
charge $S_i=+1$ for all $i$. The masses and CS parameters are set to 
$-N/2$. The FI parameter is $\tilde{\zeta}$.

\paragraph{}
It is well known that the classical Higgs branch of Theory B 
is ${\bf CP}^{N-1}$ of K\"ahler class $\tilde{\zeta}$ \cite{wit}, 
so we concentrate on the Coulomb branch of Theory A. Notice firstly 
that the charges $R_i^a$ and $S_i$ satisfy \eqn{theone}. 
Following the prescription above, we set $C_i=m$ for $i=1,N$ and 
$C_i=0$ for $i=2,\cdots,N-1$. The hyperplanes $D_i$ defined in 
\eqn{dual} are then given by 
$Y_i-Y_{i-1}=m$, where it is taken that $Y_0\equiv Y_N\equiv 0$. The vertices 
of $\nabla$ are given by the intersection points of any $N-1$ of 
the $N$ hyperplanes, 
$V^a_i=D_1\cap\cdots\cap\hat{D}_i\cap\cdots\cap D_N$, where the 
hat denotes omission of the $i^{\rm th}$ hyperplane. We thus  
find that $V^a_i=am$ for $i>a$ and $V_i^a=(a-N)m$ for $i\leq a$. 
Notice that, for $m\in {\bf Z}$, all $V^a_i\in{\bf Z}$ and 
$\nabla$ is a reflexive integral polytope as required to 
define a mirror toric variety. Moreover, as $\sum_{i=1}^NS_iV_i^a=0$ 
the vertices of $\nabla$ define another ${\bf CP}^{N-1}$. This is 
simply the statement that the complex projective space is self-mirror. 
In order to see that the Coulomb branch is defined in terms of 
$\nabla$, we examine the effective CS coupling,
\be
\keff_{ab}=\kappa_{ab}+\ft12\sum_{i=1}^NR^a_iR^b_i\ \sign (M_i)
\nn\ee
where $M_1=\phi_1-m$, $M_i=\phi_i-\phi_{i-1}$ for $i=2,\cdots,N-1$ 
and $M_N=-\phi_{N-1}-m$. 
Identifying $Y_a=\phi_a$, we see 
that the equation for the hyperplane $D_i$  is simply 
$M_i=0$ and, with the bare CS coupling given by equation \eqn{itskagain}    
to be $\kappa_{ab}=\ft12\sum_iR_i^aR_i^b$, we find that $\keff_{ab}=0$ 
for $\sign M_i =-1$ or, alternately, 
\be
-m<\phi_{N-1}<\phi_{N-2}<\cdots <\phi_1<m
\nn\ee
which is indeed the polytope $\nabla$. Further, on each boundary 
$M_i=0$, the scalar $q_i$ becomes massless and the corresponding 
cycle arising from shifts in $\sum_a R^a_i\sigma_a$ must shrink.
Choosing the $\zeta_a$ such that $\zeff_a=0$, the Coulomb branch 
is then given as a toric variety to be ${\bf CP}^{N-1}$. 
Notice that in the case of $N=2$, the above theory differs from the 
self-mirror theory introduced in the previous section. 

Although in the above example the moduli space of vacua is compact, 
more generally this will not be the case. In fact, compactness is 
assured only if the fan $\Delta$ spans ${\bf Z}^r$. In particular, in 
the case of a toric variety which obeys the Calabi-Yau condition 
$\sum_{i=1}^N{R_i^u}=0$ for all $u$, the Coulomb branch is always 
non-compact. 

\section*{Appendix: A Brane Configuration}

In this appendix we show how the vacuum structure of the self-mirror 
theory exhibited in Section 3 and, in particular, the compactness of 
the Coulomb branch can be rederived from a brane realisation of this 
theory. The relevant brane configuration is the T-dual of that 
described in \cite{hanhori}. Set-ups identical to the one discussed
here were also considered in 
\cite{bhkk}. We work in units $\alpha^\prime=1$ and set $g_{\rm st}=1$. 

The configuration of interest involves a D3-brane suspended between 
various 5-brane webs in IIB string theory. The first of these webs 
consists of an NS5-brane spanning worldvolume directions 012345, two 
D5-branes spanning worldvolume directions 012346 and two $(1,1)$ 
5-branes in directions 01234(56), where the direction in $(X^5,X^6)$-plane 
is $X^5=\pm X^6$. The final configuration is shown 
in figure 1 and is positioned at $X^7=X^8=X^9=0$. The two D5-branes 
are located at $X^5=\pm m$ while, when $m=0$, the NS5-brane 
is located at $X^6=0$. The position of this brane for general $m$ 
will be described below.

The second 5-brane web is much simpler, consisting of a 
single NS5-brane spanning 012589, which is traditionally called the 
NS$^\prime$5-brane. It is positioned at $X^3=X^4=0$, $X^7=1/e^2$ 
and $X^6=\zeta$. Finally a D3-brane is suspended between 
the web and NS$^\prime$5-brane, spanning 0127. It is of finite 
length in the $X^7$ direction and is positioned at $X^5=\phi$. 

The low-energy dynamics of the D3-brane in this configuration
are governed by the three-dimensional gauge theory of Section 3, 
with the dictionary between brane and field theory moduli explicitly 
given above. Before examining the vacuum 
structure, we must first describe the zero mode of the 5-brane 
web corresponding to changing $m$. This is denoted by the dotted 
lines in Figure 1. Notice that, 
fixing the position of the $(1,1)$ 5-branes at infinity, the 
position of the NS5-brane is given by $\zeta=-m$. 

\begin{figure}
\begin{center}
\epsfxsize=3.0in\leavevmode\epsfbox{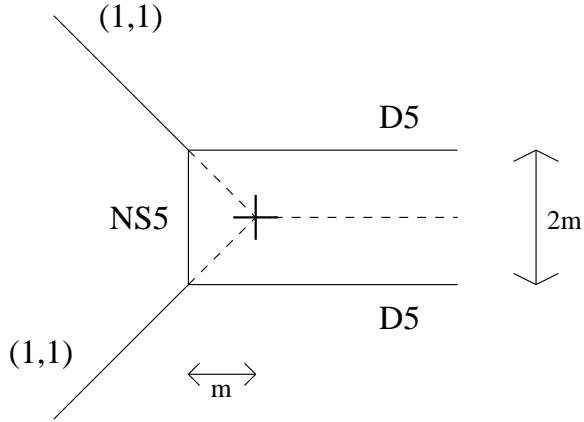}
\end{center}
\caption{The 5-brane web. The cross denotes the origin of the $(X^5,X^6)$ 
plane. The dotted lines represent the zero mode of the web.}
\label{fig1}
\end{figure}

Let us now discuss the vacuum structure of the theory. The 
numbering here coincides with that of section 2. 
Each case is illustrated with a diagram in Fig 2 in which we 
draw only the 5-brane web, with the position of the D3-brane (which 
also encodes the $X^6$ position of the NS$^\prime$5-brane) marked by 
a solid dot. 

i) $m=\zeta=0$: The set up is shown in Figure 2i). It is clear that 
the D3-brane has a unique vacuum state at $\phi=0$.

ii) $m=0$, $\zeta>0$: This is shown in Figure 2ii). The D3-brane ends on
the two coincident D5-branes located at $\phi=0$. 
This is the Higgs branch of the gauge theory which, in the analogous
two-dimensional theory, was 
argued in \cite{hanhori} to contain a copy of ${\bf CP}^1$ that is 
hard to see from the brane picture. 

iii) $m>0$, $\zeta=-m$: It is clear from the brane diagram, shown 
in Figure 2iii), that the D3-brane must end on the NS5-brane and is 
is therefore restricted to the interval $|\phi|<m$. This is the
Coulomb branch of the gauge theory.

iv) $m>0$, $\zeta>-m$: There are two isolated vacua where the D3-brane
ends on one or the other D5-brane as shown in Figure 2iv). 
These are located at $\phi=\pm m$ in agreement with the field theory 
analysis of Section 2. 

v) $m \geq 0$, $\zeta<-m$. In this case the D3-brane must end on one 
of the two dyonic fivebranes as shown in Figure 2v). The corresponding 
vacua are located at $\phi=\pm \zeta$.     

\begin{figure}
\begin{center}
\epsfxsize=5.0in\leavevmode\epsfbox{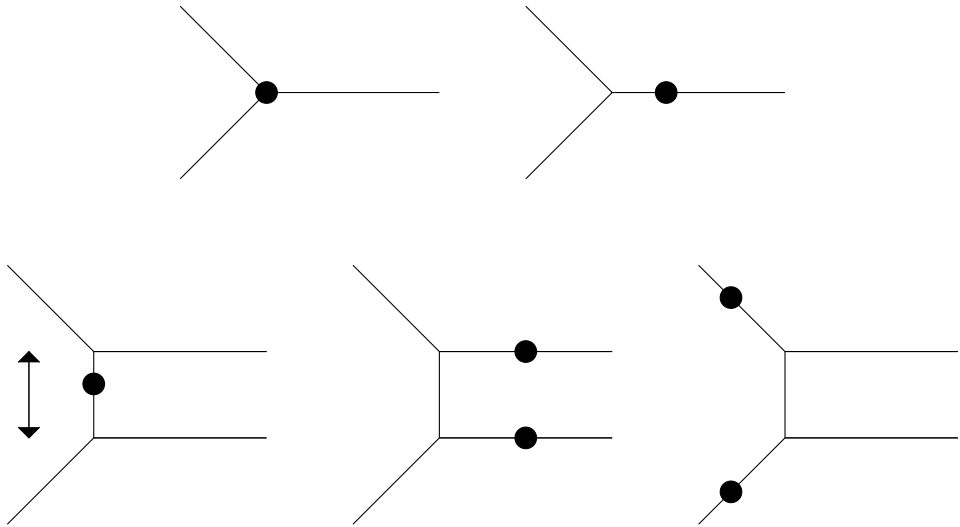}
\end{center}
\caption{ The 5-brane web for various parameters. 
The dot denotes the position of the D3-brane. The top two diagrams 
are, from left to right, 2i) and  2ii) and the bottom three 2iii),
2iv) and 2v). In diagram 2iii), the 
D3-brane is restricted to lie on the interval as shown by the arrow.}
\label{fig2}
\end{figure}

Notice that the 5-brane web encodes information about quantum 
effects of the gauge theory, in particular the contribution to the 
renormalised FI parameter arising from weakly gauged global symmetries 
(the last term in \eqn{zeff}). Moreover, it was argued in \cite{ohta,bhkk} that 
when the D3-brane is suspended between 5-branes of different kinds 
CS terms appear on its world-volume theory. This is 
again in agreement with the field theory dynamics whereby, for instance, 
in the Coulomb phase described above 
the D3-brane cannot move outside the interval 
due to dynamically generated CS terms. In fact, one can further study 
the theory of Section 3  with a bare CS coupling $\kappa=\pm 1$ 
by replacing the NS$^\prime$5-brane with a $(1,1)^\prime$-5-brane. 
It is simple to see that, for $\zeff =0$ and $m\neq 0$, the D3-brane 
may lie anywhere on one of the $(1,1)$-5-branes of the web, parametrised 
by the half-line. A short calculation confirms that this is in agreement 
with the field theory result.

The similarity between the 5-brane webs and toric skeletons was 
pointed out by Vafa and Leung \cite{vl}. In the case of an 
NS$^\prime$5-brane above, we saw that, on the Coulomb branch, the 
D3-brane was restricted to lie on the interval of the NS5-brane: 
this interval is the toric skeleton for ${\bf CP}^1$. Similarly, 
in the case of a $(1,1)^\prime$-5-brane, the D3-brane is 
constrained to lie on the half-line of the $(1,1)$-5-brane: 
this is the toric skeleton for the complex plane. In the field theory, 
using the results of Section 3 and \cite{ahiss}, one may see that the 
size of the ${\bf S}^1$ is zero at the origin of 
the half-line while asympotically, where one may trust the one-loop 
calculation, it grows linearly, in agreement with the proposal that 
the Coulomb branch is indeed the complex plane.

\subsubsection*{Acknowledgements}

We would like to thank Heather Russell for discussions on toric 
geometry. DT would further like to thank the University of Washington, 
Seattle where this work was initiated, as well as the Tata Institute of  
Fundamental Research, Mumbai and the Mehta Research Institute, Allahabad, 
for their kind hospitality. DT is supported by an EPSRC fellowship. 
ND is supported by a PPARC ARF. The authors acknowledge support from 
TMR network grant FMRX-CT96-0012.


\begin{thebibliography}{99}
\bibitem{is} K. Intriligator and N. Seiberg, ``{\em Mirror Symmetry 
in Three Dimensional Gauge Theories}'', hep-th/9607207, Phys. Lett. 
{\bf B387} (1996) 513.
\bibitem{hw} Hanany and Witten, ``{\em Type IIB Superstrings, BPS 
Monopoles, And Three-Dimensional Gauge Dynamics}'', hep-th/9611230, 
Nucl. Phys. {\bf B492} (1997) 152. \\
M. Porrati and A. Zaffaroni, ``{\em M-Theory Origin of Mirror Symmetry 
in Three Dimensional Gauge Theories}'', hep-th/9611201, 
Nucl. Phys. {\bf B490} (1997) 107.
\bibitem{berk} J. de Boer, K. Hori, H. Ooguri and Y. Oz, 
``{\em Mirror Symmetry in Three Dimensional Gauge Theories, Quivers and 
D-branes}'', hep-th/9611063, Nucl.Phys. {\bf B493} (1997) 101.\\
J. de Boer, K. Hori, H. Ooguri, Y. Oz and Z. Yin, ``{\em Mirror Symmetry 
in Three Dimensional Gauge Theories, $SL(2,Z)$ and D-Brane Moduli 
Spaces}'', hep-th/9612131, Nucl.Phys. {\bf B493} (1997) 148. 
\bibitem{berk2} J. de Boer, K. Hori, Y. Oz and Z. Yin, ``{\em Branes 
and Mirror Symmetry in N=2 Supersymmetric Gauge Theories inThree 
Dimensions}'', hep-th/9702154, Nucl.Phys. {\bf B502} (1997) 107.  
\bibitem{ahiss}  O. Aharony, A. Hanany, K. Intriligator, N. Seiberg and 
M.J. Strassler, ``{\em Aspects of N=2 Supersymmetric Gauge Theories 
in Three Dimensions}'', hep-th/9703110, Nucl. Phys. {\bf B499} (1997) 67. 
\bibitem{as} A. Kapustin and M.J. Strassler, ``{\em On Mirror 
Symmetry in Three-Dimensional Abelian Gauge Theories}'', hep-th/9902033, 
JHEP {\bf 04} (1999) 021.
\bibitem{ak} M. Aganagic and A. Karch ``{\em Calabi-Yau Mirror Symmetry 
as a Gauge Theory Duality}'', hep-th/9910184.
\bibitem{kog} L. Cooper, I. I. Kogan, R. J. Szabo, ``{\em Mirror Maps 
in Chern-Simons Gauge Theories}'', hep-th/9710179, Ann. Phys. {\bf
268} (1998) 61. 
\par L. Cooper, I. I. Kogan, R. J. Szabo, ``{\em Dynamical Description
of Spectral Flow in N=2 Superconformal Field Theories}'',
hep-th/9702088, Nucl. Phys. {\bf 498} (1997) 492.
\bibitem{red} A.N. Redlich, ``{\em Gauge Noninvariance and Parity Violation 
of Three-Dimensional Fermions}'', Phys. Rev. Lett. {\bf 52} (1984) 18. \\
A. N. Redlich, ``{\em Parity Violation and Gauge Non-invariance of the 
Effective Gauge Field Action in Three Dimensions}'', Phys. Rev. {\bf D29}, 
(1984) 2366.
\bibitem{wit} E. Witten, ``{\em Phases of N=2 Theories in Two Dimensions}'', 
hep-th/9301042, Nucl .Phys. {\bf B403} (1993) 159.
\bibitem{vl} N.C. Leung and C. Vafa, ``{\em Branes and Toric Geometry}'', 
hep-th/9711013, Adv. Theor. Math. Phys. {\bf 2} (1998) 91. 
\bibitem{mp} D. Morrison and M.R. Plesser, ``{\em Summing the Instantons: 
Quantum Cohomology and Mirror Symmetry in Toric Varieties}'', 
hep-th/9412236, Nucl. Phys. {\bf B440} (1995) 279.
\bibitem{kmv} S. Katz, P. Mayr and C. Vafa, ``{\em Mirror Symmetry and 
Exact Solution of 4D N=2 Gauge Theories - I}'', hep-th/9706110, Adv. Theor. 
Math. Phys. {\bf 1} (1998) 53.
\bibitem{batty} V. Batyrev, ``{\em Dual Polyhedra and Mirror Symmetry 
for Calabi-Yau Hypersurfaces in Toric Varieties}'', alg-geom/9310003, 
J. Alg. Geom. {\bf 3} (1994) 493.
\bibitem{hanhori} A. Hanany and K. Hori, ``{\em Branes and N=2 Theories 
in Two Dimensions}'', hep-th/9707192, Nucl. Phys. {\bf B513} (1998) 
119.
\bibitem{bhkk} O. Bergman, A. Hanany, A. Karch and B. Kol, ``{\em 
Branes and supersymmetry breaking in 3D gauge theories}'', 
hep-th/9908075.
\bibitem{ohta} T. Kitao, K. Ohta and N. Ohta, ``{\em Three-Dimensional 
Gauge Dynamics from Brane Configurations with (p,q)-Fivebrane}'', 
hep-th/9808111, Nucl.Phys. {\bf B539} (1999) 79.



\end{thebibliography}
\end{document}